# Do dielectric bilayer metasurfaces behave as a stack of decoupled single-layer metasurfaces?

**ALFONSO PALMIERI[1],[†], AHMED H. DORRAH[1],[†], JUN YANG[2], JAEWON OH[1], PAULO DAINESE[2],[*], AND FEDERICO CAPASSO[1],[*]**

[1]*Harvard John A. Paulson School of Engineering and Applied Sciences, Harvard University, Cambridge, Massachusetts 02138, USA*
[2]*Corning Inc., Painted Post, New York 14870, United States*
[†]*These authors contributed equally*
*DaineseP@corning.com; Capasso@seas.harvard.edu*

**Abstract:** Flat optics or metasurfaces have opened new frontiers in wavefront shaping and its applications. Polarization optics is one prominent area which has greatly benefited from the shape-birefringence of metasurfaces. However, flat optics comprising a single layer of meta-atoms can only perform a subset of polarization transformations, constrained by a symmetric Jones matrix. This limitation can be tackled using metasurfaces composed of bilayer meta-atoms but exhausting all possible combinations of geometries to build a bilayer metasurface library is a very daunting task. Consequently, bilayer metasurfaces have been widely treated as a cascade (product) of two decoupled single-layer metasurfaces. Here, we test the validity of this assumption for dielectric metasurfaces by considering a metasurface made of titanium dioxide on fused silica substrate at a design wavelength of 532 nm. We explore regions in the design space where the coupling between the top and bottom layers can be neglected, i.e., producing a far-field response which approximates that of two decoupled single-layer metasurfaces. We complement this picture with the near-field analysis to explore the underlying physics in regions where both layers are strongly coupled. We also show the generality of our analysis by applying it to silicon metasurfaces at telecom wavelengths. Our unified approach allows the designer to efficiently build a multi-layer dielectric metasurface, either in transmission or reflection, by only running one full-wave simulation for a single-layer metasurface.

## 1. Introduction

Metasurfaces or flat optics refer to subwavelength-spaced arrays of scatterers with spatially varying geometries (shape, size, and orientation) and have been widely utilized as a compact wavefront shaping tool [1–9]. Metasurfaces made of shape-birefringent nanofins have unlocked many possibilities in polarization optics ranging from vectorial structured light and holography to imaging and polarimetry [10–17]. While many wavefront shaping capabilities have been demonstrated using plasmonic metasurfaces [18–24], dielectric metasurfaces have gained much attention recently, especially for phase and polarization control, due to their relatively low loss. Dielectric form birefringent meta-atoms can be locally represented by a spatial arrangement of linearly birefringent wave-plate-like elements, each mathematically described in Cartesian coordinates by the Jones matrix formalism [15]. A general Jones matrix has four components, each with two degrees of freedom, amplitude and phase. Hence, there is a maximum of eight degrees-of-freedom, or free parameters, that can be tuned in any linear material. The more free parameters that can be varied in a Jones matrix, the more diverse the capabilities of the structure to manipulate light's polarization [10, 11, 14, 15].

The Jones matrix that can be physically realized with a single-layer dielectric metasurface is subject to being unitary and symmetric [15, 25]. This sets a fundamental constraint on the retardance and di-attenuation values (i.e., possible polarization transformations) which can be imparted on incoming light. Retardance and di-attenuation respectively refer to the relative phase and amplitude imparted on two input orthogonal polarizations. Notably, matrix symmetry is a fundamental constraint that cannot be surmounted by design. It originates from the linear shape-birefringence of dielectric metasurfaces which fails to realize circular or elliptical birefringence. Hence, a single-layer metasurface cannot be used to build a circular polarizer/retarder — its eigen polarizations must be linear. This limitation exists in any single-layer meta-atom with vertical sidewalls (regardless of its geometry) as long as it is reciprocal [26]. Surmounting this constraint requires breaking the in-plane symmetry; either using slanted or a bilayer stack of meta-atoms. Unitarity, on the other hand, is a less fundamental constraint that stems from the lossless nature of dielectric meta-atoms and the typical choice to operate off-resonance to realize higher efficiencies. Nevertheless, a unitary metasurface can still modulate both the amplitude and phase of incoming light by dumping light onto the diffraction orders (which behave as loss channels) as is standard in holography [27, 28].

By combining the propagation phase (which arises from varying the dimensions of the nanofins/meta-atoms) and geometrical phase (related to the rotation angle of the nanofin about its axis), it is possible to implement a symmetric Jones matrix with 3 independent DOF—namely, two different phase terms on the diagonal and two identical off-diagonal phase terms. With this functionality, a single metasurface can impart two independent phase [11, 29] or amplitude [30] profiles on any two input orthogonal polarizations (linear or elliptical), with the caveat that input elliptical polarizations will flip their handedness at the output. Using clever arrangements of metasurface unit cells, a Jones matrix with more DOFs can be realized. For instance, super-cell based single-layer metasurfaces can offer six DOFs for the Jones matrix, enabling complex amplitude modulation on two orthogonal polarization bases in the far field [27, 28]. Another variation of super-cell metasurfaces have recently been used to generate multiple polarization-sensitive holograms (exceeding 10 chanels) by exploiting higher diffraction orders as energy loss channels [31]. In all these works, however, one cannot freely decouple the input and output polarization states, hindered by matrix symmetry. To tackle this limitation, and access all 8 DOFs of the Jones matrix, the in-plane symmetry must again be broken by constructing a multi-layer system [32]. In principle, a bilayer metasurface can impart arbitrary and independent amplitude and phase control on any set of two orthogonal polarizations, while completely decoupling the input and output polarization states [33]. Multi-layer metasurfaces also suggest new ways for achieving lossless polarization transformations [34] as well as non intuitive ways for wavefront shaping by utilizing more general configurations of the Pancharatnam-Berry phase [35].

With recent advances in nanofabrication, multi-layer metasurfaces have now become feasible. For instance, by cascading two single-layer metasurfaces made of silicon at design wavelength of 808 nm, each with six DOFs in their Jones matrix, a spatially varying Jones matrix with full parameters of eight DOFs has been realized [36]. The latter has been utilized in polarization-selective holography with 16 different channels. In addition, compound meta-optics made of $\alpha$-silicon in the near infrared (in combination with inverse design) have been utilized in spatial mode multiplexing, optical mode conversion, and vectorial holography with very high diffraction efficiencies [37, 38]. On the other hand, the design of compound metasurfaces (made of either bilayer or cascaded meta-atoms) is computationally intensive compared to single-layer devices. For example, building an exhaustive bilayer library is not a straight-forward process due to the huge size of the parameter space which requires varying the nanofins dimensions ($D_x$ and $D_y$, assuming a rectangular geometry) for both the top and bottom layers while performing the simulation at different angular rotations between the two. Furthermore, assuming a library is built, mapping a target 2-by-2 Jones matrix profile, pixel by pixel, to this massive library is a challenging and time consuming task. Accordingly, bilayer dielectric metasurfaces have been widely treated as a stack of two decoupled metasurfaces. A rigorous validation of this general assumption, however, has not been presented in the literature. To address this gap, here we use full-wave simulations to build a bilayer metasurface library and we study the coupling between the top and bottom nanofins by varying the full design space and analyzing the far-field and near-field responses. Our aim is to provide a systematic recipe that allows the designer to narrow down regions in the parameter space where the top and bottom nanofins are effectively decoupled while avoiding the geometries that suffer from strong bilayer coupling. This viewpoint simplifies the construction of a bilayer metasurface as it only requires one full-wave simulation for a single-layer nanofin in transmission. Using simple Jones calculus, the output response of the former can then be cascaded (for e.g., using matrix multiplication) to build a multi-layer structure. We perform our analysis for bilayer metasurfaces operating in transmission and reflection. Although we studied $TiO_2$ at the visible wavelength range because of the limited literature on bilayer metasurfaces in that regime, our analysis provides a holistic guideline to the metasurface community which can also be applied to various wavelengths and material platforms as will be shown.

## 2. Analysis

### 2.1. Operating in transmission

We start by creating a library for a single-layer metasurfaces using the finite-difference time-domain (FDTD) and represent its response in terms of a Jones Matrix [11]. A model of the simulated structure is shown in Figure 1(a). It is made of a titanium dioxide ($TiO_2$) rectangular nanofin on top of a fused silica substrate and can impart two different phase delays along its major and minor axes; hence the shape-birefringence. The boundary condition applied at the edges of the simulation box is the Periodic Boundary Condition (PBC) which emulates an infinitely periodic array of the same rectangular nanofins. By sweeping the nanofin dimensions and recording the phase and transmission response in the far-field, a single-layer metasurface library can be built (as depicted in Supplementary Figure 1). The response of each nanofin to *x*- and *y*-polarized light can then be mathematically cast in a 2-by-2 Jones matrix.

As a next step, we extend our analysis to the case of a bilayer metasurface. We explore the possibility of computing the Jones matrix of the bilayer as the product of two decoupled (single-layer) Jones matrices. This requires that the coupling between the two nanofins is negligible regardless of the dimensions of the two nanofins and their relative rotation. To test this assumption, we compare the results of the FDTD simulations of the bilyer with the "analytical" results obtained from the product of two single-layer metasurfaces. The model of the simulated structure is shown in Fig. 1(b). Initially, we fixed the dimension of the bottom nanofin at 134 nm x 202 nm, an arbitrarily chosen geometry which emulates a quarter-wave plate. The Jones matrix of this geometry was then extracted from the single-layer library reported in Supplementary Figure 1. Afterwards, we performed a parameter sweep for the dimensions of the top nanofin without introducing a rotation angle between the nanofins; i.e., keeping the angular orientation of both nanofins at 0°. This simulation helps us verify if the two layers can be considered decoupled for all the geometries. If that is true, then the Jones matrix of the bilayer could be expressed as:

$$J_{\text{bilayer}} = J_{\text{top}} \cdot J_{\text{bottom}} = \begin{bmatrix} e^{i\phi_{x,1}} & 0 \\ 0 & e^{i\phi_{y,1}} \end{bmatrix} \cdot \begin{bmatrix} e^{i\phi_{x,2}} & 0 \\ 0 & e^{i\phi_{y,2}} \end{bmatrix}. \tag{1}$$

The transmission and phase responses for this bilayer structure under *x*-polarized illumination are depicted in Figs. 1(c-d), respectively. From these plots, one can observe transmission dips which can be attributed to resonances. Here, we refer to all geometries whose transmission is lower than 80 % to be in the regime of resonance. In Fig. 1(e), we plot the error between the phase response (from the full-wave simulation above) and the analytical phase response calculated using Eq. (1). Since the rotation angle between the two nanofins is zero ($\theta = 0$) the Jones matrix that describes the structure is diagonal. Hence, in the plots we only show the results related to the element $J_{11}$. The error plots for element $J_{22}$ (under incident y-polarization) are similar due to the rectangular geometry and are thus not included. The average absolute phase error here is less than 3°. For most of the geometries, the simulation results coincide with the analytical ones. The few geometries that exhibits a larger error (only 2% of the geometries have an error higher than 10°) are the same ones that correspond to the resonance lines in the top right of Fig. 1(d). Therefore, for these geometries and others with large phase error, a full-wave simulation of the bilayer structure is needed to accurately capture its Jones matrix. However, given that full 0-2$\pi$ phase coverage can be achieved with enough geometries away from resonance, these resonant elements can simply be filtered out from the library. In the next section, we will provide a more in depth investigation of these resonances (and their types) and demonstrate cases in which the coupling effects in a bilayer meta-atom due to resonance can be neglected.

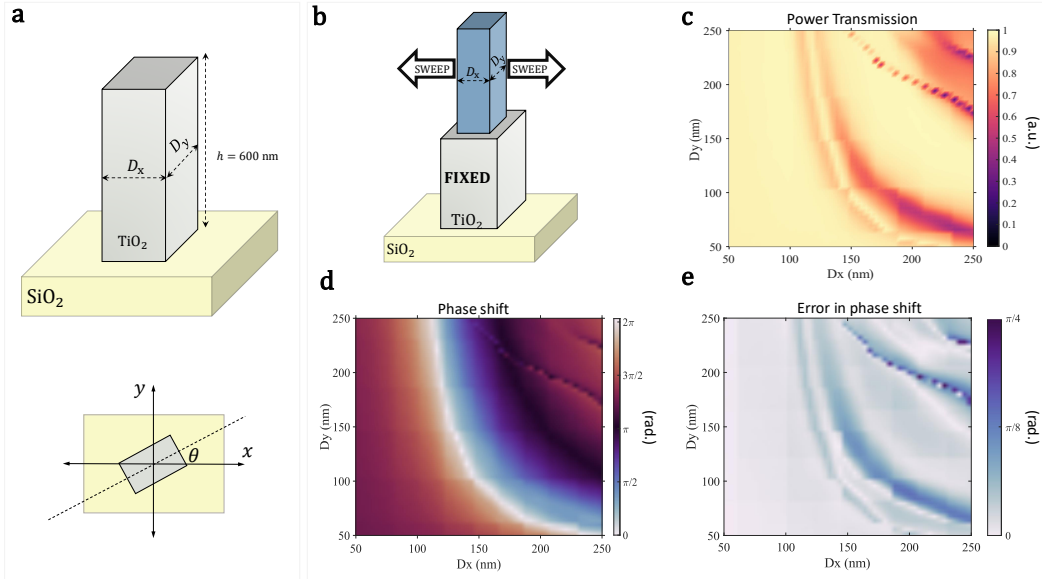

Figure 1. Bilayer dielectric metasurface with fixed bottom nanofin, operating in transmission. **(a)** Model of the unit cell of a transmissive single-layer metasurface. **(b)** Model of the unit cell of the simulated bilayer. The dimensions of the bottom nanofin are fixed and set to be 134 nm × 202 nm. The dimensions of the top nanofin are swept from 50 to 250 nm. The Power transmission and phase response of the bilayer metasurface are shown in **(c)** and **(d)**, respectively. **(e)** Absolute error in phase shift between the simulation results shown in (d) and the results analytically obtained from the matrix product of two cascaded single-layer metasurfaces.

To study the case of a non-diagonal Jones matrix, we introduce a relative rotation between the top and bottom nanofins. In Supplementary Note 2, we considered two cases of bilayer structures made of identical top and bottom nanofins and we rotated the top nanofin at increments of 15° to study the effect of angular orientation on coupling. The two bilayer structures are made of identical nanofin with dimensions 134 nm × 202 nm and 114 nm × 154 nm, respectively. For each of the two structures, we tabulated the amplitude and phase error for each element in their corresponding Jones matrices. Our analysis suggests an average phase error — between FDTD simulations and analytical calculation of Eq. (1) — on the order of 5°. The phase error plots are shown in Supplementary Figure 2 for one of the two geometries confirming that 0 to $2\pi$ phase coverage can be achieved with acceptable errors.

As a next step, we extend our analysis by performing a full sweep for the dimensions of both the top and bottom nanofins. We then examine the effect of bilayer coupling by comparing these FDTD simulation results with the analytical calculation of Eq. (1), as before. Figures 2(b-c) depict the phase shift and power transmission obtained from FDTD. Each figure exhibits a grid that is made of 121 subplots (cells). Each cell represents one parameter sweep where the top nanofin has the dimensions reported on the horizontal/vertical axes while the dimensions of the bottom nanofin are swept from 50 nm to 250 nm. Interestingly, the response of each subplot mimics the behaviour of the entire plot, as if it is of fractal nature. This behavior can be reconciled with sampling theorem since the figure is compiled from a Jones matrix product, between the top and bottom nanofin, that is reminiscent of a convolution between their respective phase plots. To quantify the bilayer coupling effects, we plot the phase and transmission errors (with respect to the analytical prediction of Eq. (1)) as shown in Figs. 2(d-e), respectively. The axes of the two plots can be interpreted the same as described above. This analysis confirms the possibility of expressing the Jones matrix of a bilayer dielectric metasurface as the product of the single-layer Jones matrices over most of the geometries included in the considered parameter space. The average phase error in this case is 3.19°.

Therefore, for a large number of geometries, the hypothesis that the two layers can be treated as decoupled is a valid one. Notably, the phase redundancy afforded by the metasurface library offers a sufficiently large number of decoupled geometries with full 0 to $2\pi$ phase coverage. As a rule of thumb, the geometries that exhibit larger errors fall within two main categories: a) structures in which one of the two nanofins is operating near resonance, and b) structures with significant reflections where the top nanofin is much larger than the bottom one. The latter can specially be inferred by examining the top right part of Fig. 2(c). This can be attributed to the multiple (Fabry-Perot like) reflections that occur between the substrate and the lower base of the top nanofin. To gain more insights into the bilayercoupling, we consider different geometries and perform a near-field analysis to examine the cases with large errors. The underlying physics of this problem is discussed next.

### 2.2. Near-field analysis

In this section, we complement the far-field Jones matrix calculations presented before with a near-field analysis. Our goal is: a) to evaluate the accuracy of our Jones matrix analysis in predicting the response of the bilayer structure, and b) to gain more insights into cases where the far-field response obtained from FDTD deviates from the Jones matrix analysis. Towards these aims, we adopted the scattering matrix (S-matrix) and transmission matrix (T-matrix) approaches while calculating the fields using the eigenmode expansion (EME) method [39–42]. Since we are interested in understanding the bilayer coupling in a transmissive metasurface, our ports are defined at the input of the bottom nanofin and at the output of the top nanofin, respectively.

To assess the effects of guided and evanescent modes as well as back reflections, separately, we adopt four different calculations: a) The full S-martrix which captures the contributions of guided, evanescent, and back reflected fields (hence, it is the most accurate calculation or ground truth). b) The full T-matrix is similar to (a) but without recording the back reflections; only forward propagating guided and evanescent fields. (c) The (2 × 2) S-matrix which only considers the fundamental propagating mode and back reflections while ignoring the evanescent fields. (d) The $(2 \times 2)$ T-matrix which only captures the forward propagating modes without recording neither evanescent fields nor back reflections. Hence, by definition, the $(2 \times 2)$ T-matrix calculation coincides with our previous far-field Jones matrix analysis. Comparing between these four calculations will help quantify the effects of bilayer coupling (which can be inferred from the strength of evanescent fields) and back reflections for various meta-atom geometries. To further examine the effects of evanescent coupling and impedance mismatch, we introduce a small air gap between the top and bottom nanofins. By varying the gap size between the two nanofins and recording the amplitude variation one can identify the regimes where bilayer meta-atoms no longer behave as a stack of two decoupled single-layer metasurfaces.

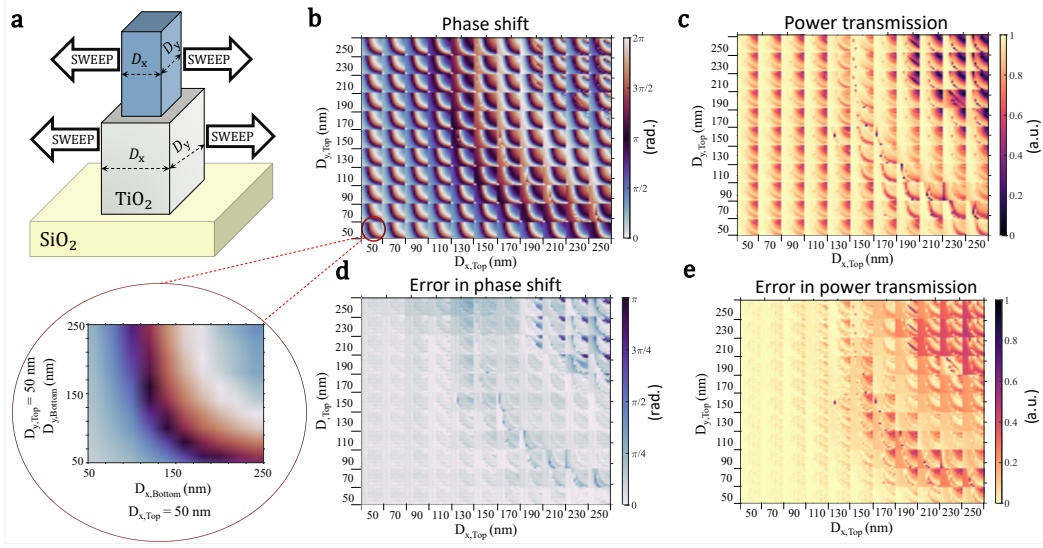

Figure 2. Bilayer dielectric metasurface with top and bottom nanofins of variable dimensions, operating in transmission. **(a)** Model of simulated bilayer metasurface: the arrows refer to sweeping the dimensions of top and bottom nanofins. The phase shift and power transmission of the structure in (a) is recorded using FDTD simulation and is shown in **(b)** and **(c)**, respectively. Each cell in these grids refers to a separate simulation in which the bottom nanofin's dimensions are swept while fixing the dimensions of the top nanofin. The top nanofin's dimensions are depicted on both axes. **(d)** Error in phase shift between the simulation results shown in (b) and the results analytically obtained from the single-layer library. **(e)** Error in power transmission between the simulation results shown in (c) and the results analytically obtained from the single-layer library.

We consider four main categories of bilayer dielectric meta-atom geometries under x- and y-polarizations, respectively. Figure 3(a) depicts the first case: a bilayer meta-atom composed of two off-resonance nanofins[*] with a larger nanofin at the bottom. By looking at the amplitude response under the input polarizations, $E_x$ (11) and $E_y$ (00), one can observe that the S-matrices transmission oscillate only slightly around a mean value that matches the T-matrices result. Here, the deviation in amplitude between the S- and T-matrices calculations is on the order of 0.01%. This implies that the $(2 \times 2)$ T-matrix (and so the Jones matrix) correctly describe the nanofin dynamics accurately; the evanescent coupling and back reflections in this type of geometry are small enough to be ignored.

---

[*]In our analysis, a resonant nanofin refers to any geometry with less than 89% transmission. This description may include nanofins which are not strictly at resonance (but rather near resonance) but is justified here as it ensures a more conservative choice of library elements.

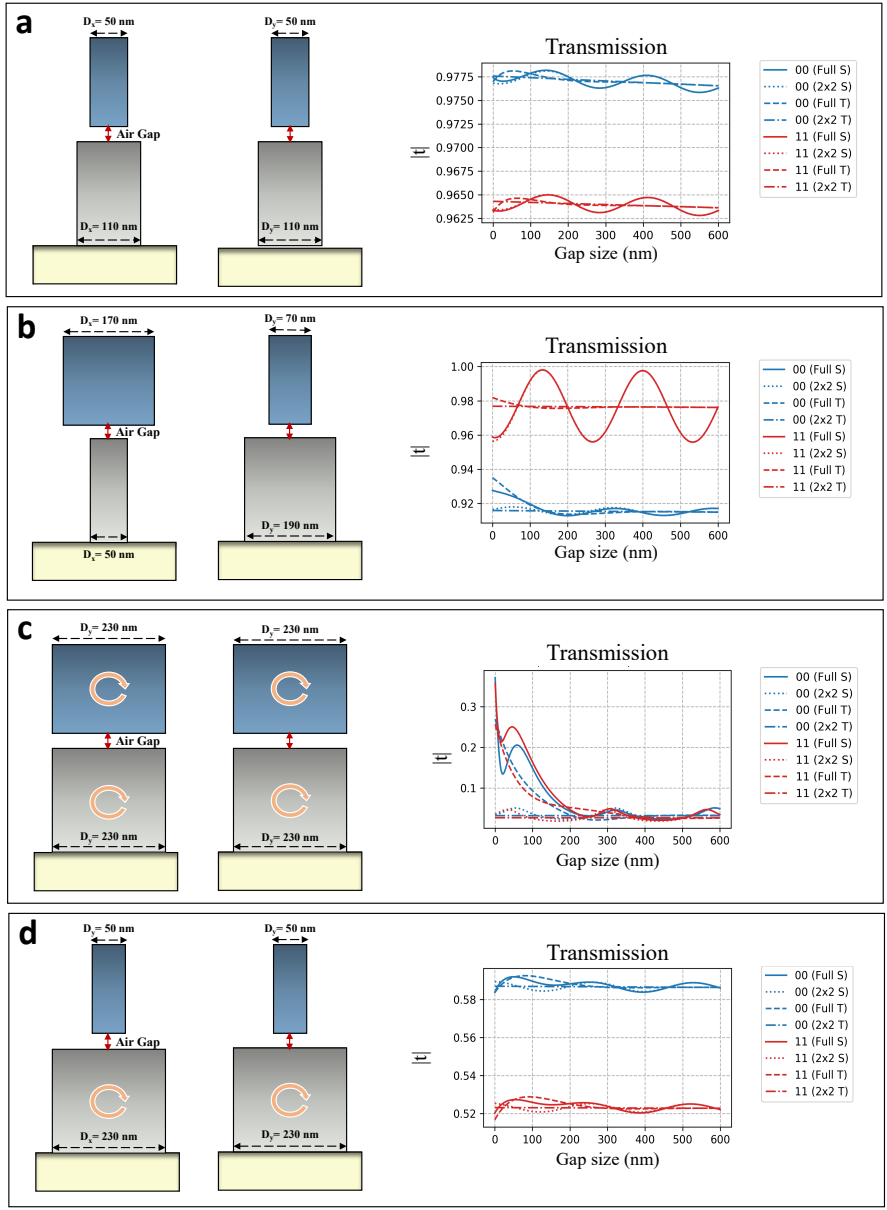

Figure 3. Scattering and transmission matrix analysis for transmissive bilayer metasurfaces with aligned top and bottom nanofins and variable air gap in between. Four cases are considered: (a) Two nanofins operating off-resonance. The top nanofin is smaller than the bottom one. The transmission amplitude response from the S- and T-matrices under two orthogonal polarizations, $E_x$ (11, blue curve) and $E_y$ (00, red curve), is plotted. (b) A birefringent bilayer meta-atom with different dimensions along x and y. The two nanofins are off-resonance and the corresponding amplitude response under x- and y-polarizations is shown. (c) Bilayer meta-atom composed of two identical nanofins operating at resonance. (d) Only the bottom nanofin of the bilayer meta-atom is at resonance and is larger than the top nanofin.

Figure 3(b) represents the second case which is anisotropic: the two nanofins have different dimensions along the x and y directions but neither of the two is at resonance. The top nanofin is larger along the x direction ($D_{x,top} > D_{x,bottom}$) and is smaller along y ($D_{y,bottom} > D_{y,top}$). Hence, under x-polarized illumination, light will be reflected due to the size mismatch between the two nanofins. In this case, a Fabry-Perot like cavity will be created between the substrate and the top nanofin. As these back reflections are not captured by the T-matrices, a large deviation in the S- and T- matrix amplitude response is observed. On the other hand, when the same nanofin is illuminated by y-polarized light (blue curves), the amplitude response evaluated by the T-matrices and S-matrices are in good agreement; the back reflections and evanescent coupling in this geometry is minimal. These observations can be reconciled with waveguide theory. Both geometries involve impedance mismatch between the top and bottom nanofins (i.e., the modes are characterized by two different effective indices). However, since the small nanofin will have a smaller effective index (close to the cladding material—air), its placement above the large nanofin is already captured by the Jones matrix of the large nanofin. This is not the case when the small nanofin is terminated by the large one on top. The impedance mismatch in the latter is more significance and is thus not fully captured by the Jones matrix of the small nanofin.

Figure 3(c) shows the case of two identical nanofins operating at resonance. In this case it is not expected that back reflection between the top nanofin and the substrate can occur (given the agreement in cross sectional area) and indeed the discrepancy between the full S-matrix and full T-matrix is insignificant. However, the large discrepancy between the full S- and full T-matrices versus the ($2 \times 2$) S- and ($2 \times 2$)T-matrices suggest that the higher order evanescent modes play a fundamental role in the amplitude response. Here, the evanescent field coupling between the two nanofins is very significant (due to the operation at resonance) and, hence, cannot be neglected. Additional categories of nanofins that feature back reflections and that exhibit coupling through evanescent fields are shown in Supplementary Figure 3. The former is typically observed when the top nanofin is larger in size whereas the latter occurs when at least one of the two nanofins is at resonance.

Figure 3(d) highlights another case of resonant geometries. Here, only the bottom nanofin is at resonance whereas the top nanofin is of much smaller dimensions. In this case, due to the small spatial overlap between the two nanofins, the evanescent coupling is not significant. The smaller dimensions of the top nanofin also suggest that the back reflections are minimal. These two behaviors—i.e., weak evanescent coupling and reflections—are indeed suggested by the response of the ($2 \times 2$) T-matrix which captures the amplitude transmission fairly accurately. Hence, the full S-matrix and full T-matrix are in close agreement.

In short, our near-field analysis confirms that (under some circumstances) the S- and T-matrices responses coincide. Cases with disagreement correspond to geometries that involve large back reflections (small spatial overlap between top and bottom nanofins) or geometries that lie close to resonance. By avoiding these regions in the parameter space, one can populate a set of bilayer meta-atoms which effectively behave as a stack of two decoupled single-layer metasurfaces. Thanks to the redundancy afforded by the metasurface library such a task is possible to achieve. Figure 4(a) depicts a schematic which visualizes the nanofin categories. The set of allowable (decoupled) nanofin geometries are represented in the first row (green zone). After filtering out all the geometries with bilayer coupling, the remaining meta-atoms (around 50% of the parameter space) still densely span from 0 to $2\pi$ phase shift as shown in Fig. 4(b).

Note that thus far we have chosen a specific material, namely titanium dioxide, as a platform for conducting our analysis. Nevertheless, we expect the physical dynamics associated with evanescent coupling and back reflection, to be universal across other dielectric metasurface libraries using different material platforms or design wavelengths. For instance, silicon nitride (SiN) is another widely used material in the visible, albeit with less index contrast [43]. Therefore, due to the less mode confinement, the coupling strength can be more significant causing all resonant geometries to be considered coupled (including for e.g., the category in Fig. 4(a)-iii. Other effects, such as back reflections, should preserve their behavior. In addition, other material platforms also exist for the near infrared and telecom regimes including, for e.g., silicon. To validate the generality of our analysis, we study bilayer dielectric metasurfaces based on silicon, at a design wavelength of 1550 nm, and show that the coupling effects are governed by the same physical dynamics. We summarize the results of this analysis in Supplementary Section 7.

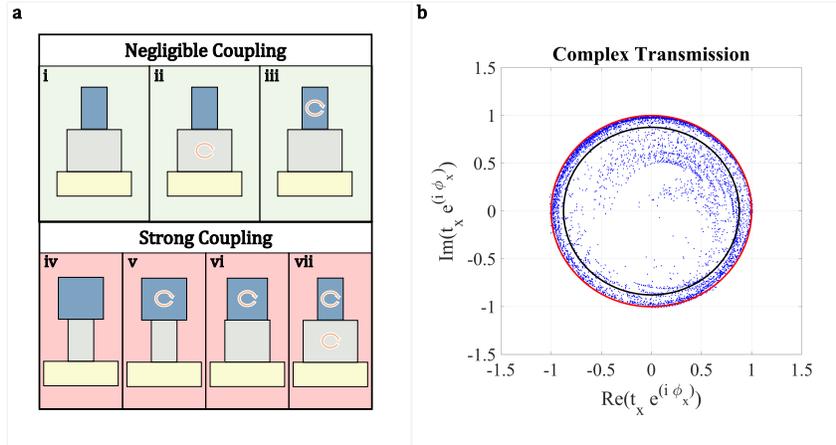

Figure 4. **(a)** Schematic of different categories of bilayer dielectric meta-atoms. The first row corresponds to the geometries for which the coupling between the nanofins is negligible: **(i)** neither the bottom nor the top nanofins are at resonance and bottom nanofin is larger than the top, **(ii)** only the bottom nanofin is at resonance and is larger than the top, **(iii)** only the top nanofin is at resonance but the bottom nanofin is much larger than the top. The second row depicts the cases for which the two nanofins in the bilayer are strongly coupled: **(iv)** neither the bottom nor the top nanofins are at resonance and the top nanofin is larger than the bottom, **(v)** only the top nanofin is at resonance and larger than the bottom, **(vi)** only the top nanofin is at resonance and slightly smaller than the bottom, **(vii)** both nanofins are at resonance. These cases are detailed more fully in Supplementary Section 4. **(b)** Complex transmission of the dashed region in (a). The electric field amplitude $t_x e^{i\phi_x}$ is plotted on the complex plane demonstrating the 0-2$\pi$ phase coverage afforded by the considered geometries while maintaining low loss. The red circle is the unit circle. The black circle is the average of $t_x$ over all considered geometries.

## 2.3. Operating in reflection

Thus far we have shown that we can evaluate the Jones matrix of a transmissive bilayer metasurface starting from the Jones matrix of a single-layer metasurface under some constraints. This allows the designer to build a bilayer metasurface by only utilizing the single-layer metasurface library, thereby simplifying the design process. In this section, we investigate the validity of this assumption for reflective dielectric metasurfaces. We examine if a birefringent metasurface operating in reflection can be expressed as the product of four Jones matrix (each describing a single-layer nanofin). In this case, the product of one unit cell (pixel) on the bilayer metasurface, assuming no rotation, is given by:

$$J_{\text{bilayer}} = J_{\text{top}} \cdot J_{\text{bottom}} \cdot J_{\text{mirror}} \cdot J_{\text{bottom}} \cdot J_{\text{top}} = \begin{bmatrix} e^{i\phi_{x,\text{top}}} & 0 \\ 0 & e^{i\phi_{y,\text{top}}} \end{bmatrix} \cdot \begin{bmatrix} e^{i\phi_{x,\text{bottom}}} & 0 \\ 0 & e^{i\phi_{y,\text{bottom}}} \end{bmatrix} \cdot \begin{bmatrix} 1 & 0 \\ 0 & -1 \end{bmatrix} \cdot \begin{bmatrix} e^{i\phi_{x,\text{bottom}}} & 0 \\ 0 & e^{i\phi_{y,\text{bottom}}} \end{bmatrix} \cdot \begin{bmatrix} e^{i\phi_{x,\text{top}}} & 0 \\ 0 & e^{i\phi_{y,\text{top}}} \end{bmatrix}. \quad (2)$$

Equation (2) describes the path that light makes when impinging on the metasurface as it passes through the two nanofins, gets reflected by the mirror, before traversing the two nanofins again, in the reverse order. To test the validity of Eq. (2), we start by simulating a unit cell consisting of a single-layer in reflection and then we build on it by considering the full bilayer metasurface in reflection. Following the same approach used for a transmissive metasurface, we consider a single-layer reflective metasurface first to verify if, in presence of a mirror, one can express the Jones matrix of the system as the following product:

$$J_{\text{reflection}} = J_{\text{transmission}} \cdot M \cdot J_{\text{transmission}} = \begin{bmatrix} e^{i\phi_x} & 0 \\ 0 & e^{i\phi_y} \end{bmatrix} \cdot \begin{bmatrix} 1 & 0 \\ 0 & -1 \end{bmatrix} \cdot \begin{bmatrix} e^{i\phi_x} & 0 \\ 0 & e^{i\phi_y} \end{bmatrix}. \quad (3)$$

Supplementary Figure 4 shows a comparison between the phase responses of the the reflective (single-layer) metasurface obtained from FDTD and and the one calculated analytically using Eq. (2), suggesting significant discrepancy between the two. Note that placing the nanofin on top of a mirror perturbs the phase response of the structure, possibly introducing standing wave patterns (similar to terminating a waveguide with a complex load). The mirror-dielectric interface effects cannot be accurately captured by Eq. (2). Therefore, in contrast to the case of transmissive bilayer metasurfaces, where the analytical approach was fairly accurate in predicting the amplitude and phase responses, the requirements are more stringent when operating in reflection. To bypass this challenge, we introduce a simple modification to the structure by inserting a layer of silica (named "spacer") between the mirror and the nanofin, essentially matching the impedance between the two, as depicted in Fig. 5(a). To compensate for the effective index perturbation of the $TiO_2$ fin which arose from the mirror-dielectric interface, the spacer dimensions need to be optimized depending on nanofin size. The goal is obtain an overall response for the reflective structure that matches the behaviour of the transmissive nanofin as if it was in direct contact with the substrate. This allows us to build a reflective bilayer metasurface using the Jones matrix product by starting from the same single-layer metasurface library in transmission.

To optimize the spacer, we sweep its thickness between 20 nm and 165 nm, and for each thickness, we perform full sweep on the dimensions of the nanofin. For each nanofin geometry, we find the optimized spacer dimension that minimizes the phase error compared to the analytical one of Eq.

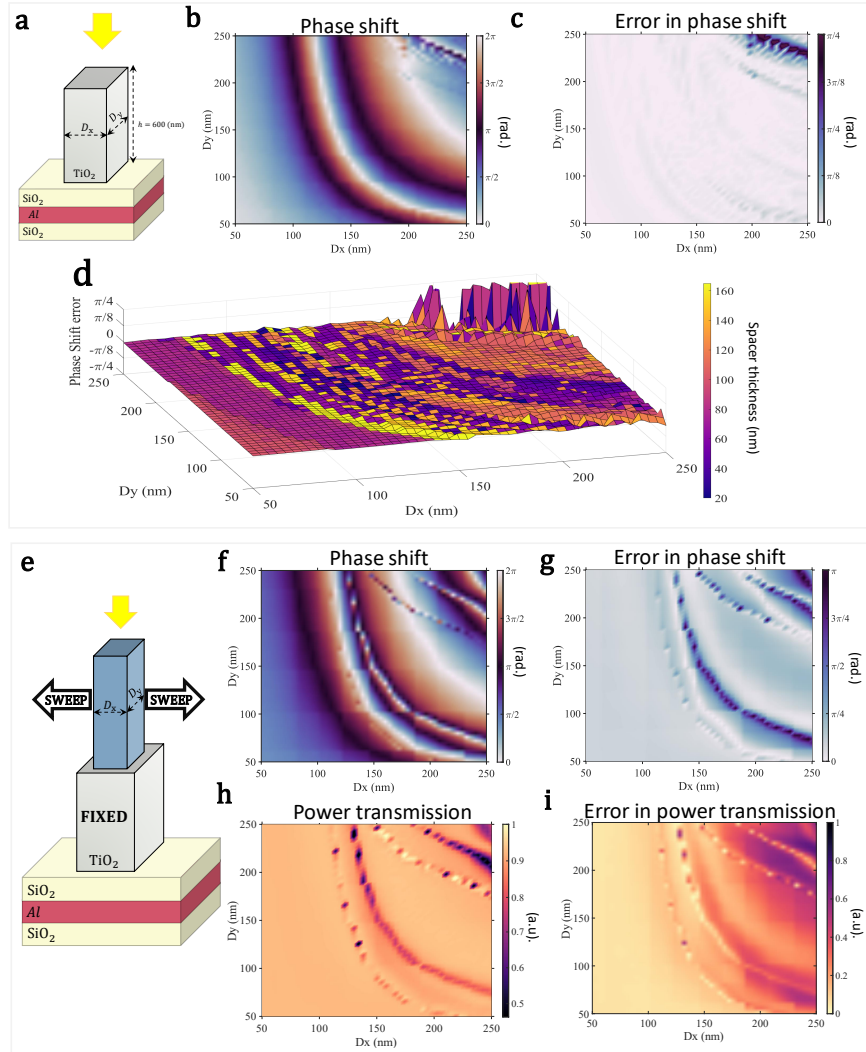

Figure 5. Operation of a reflective metasurface. (**a**) Model of the unit cell of a single-layer metasurface operating in reflection. A layer of silica has been inserted between the mirror and the nanofin. (**b**) Phase shift obtained from the FDTD simulation of the structure shown in (a). For each geometry the spacer thickness has been optimized so that the phase difference between the simulation results and the analytical product is minimized. (**c**) Difference between the simulated and analytical phase shifts for the structure shown in (a) while optimizing the spacer thickness for each geometry. (**d**) Optimum spacer dimension as a function of the nanofin geometry. The vertical axis depicts the phase error for each optimum spacer thickness. (**e**) Model of the unit cell of bilayer metasurface working in reflection. (**f**) Phase shift obtained from the FDTD simulation of the structure shown in (e). (**g**) Phase shift error for the structure shown in (e) with optimized spacer thickness for each geometry. (**h**) Power transmission obtained from FDTD simulation for the structure shown in (e). **i** Error in power transmission for the structure in (e) when the spacer thickness is optimized for each geometry.

(4). Figure 5(b) shows the phase shift obtained for each geometry when its optimized spacer thickness is selected. The error between the simulated results and the analytical ones is 0.5°, on average, provided that the spacer thickness is optimized, as depicted in Fig. 5(c). In addition, Fig. 5(d) shows the optimum spacer thickness as a function of the nanofin dimensions. Here, the vertical axis depicts the phase error between FDTD and the analytical product of Eq. (4). Therefore, it is now possible to express a single-layer reflective metasurface as the Jones matrix product of single-layer decoupled nanofins. This intermediate step is essential as it enables us to design a bilayer metasurface by only relying on a single-layer library. However, the insertion of the matching spacer obviously imposes a constraint on the design of the metasurface; as it is not possible (at least with our current fabrication techniques) to build a device with spatially varying spacer thickness. Instead, we envision the final device to be made of a bilayer meta-atom where the bottom nanofin is of fixed dimensions while only sweeping the top nanofin. A metasurface unit cell of this kind can still realize an asymmetric (yet unitary) Jones matrix, point-by-point, across the structure. By making use of super-cell metasurfaces or one can break the unitarity constrainst as well.

To design a bilayer metasurface in reflection we repeat the previous analysis performed in transmission. We fix the dimension of the bottom nanofin at 134 nm × 202 nm. This is justified because a stack of two nanofins provides 6 degrees-of-freedom, whereas an arbitrary Jones matrix requires only 4 [36]. Hence, by fixing the bottom nanofin and varying the top one, all 4 degrees-of-freedom can still be accessed. We set the spacer thickness 100 nm which is the optimized value for that selected geometry as suggested by Fig. 5(d). We performed a parameter sweep of the dimensions of the top nanofin without introducing a rotation angle between the fins. The previous analysis on the spacer optimization was needed to select the spacer dimension that optimizes the response of the fixed bottom geometry. This allow us to verify the assumption of decoupling in reflection for all the geometries. We also confirm that the spacer dimension from the single-layer is valid for the design of a bilayer metasurface in reflection; in which case the Jones matrix of the bilayer can be written as shown in Eq. (2). The results of the simulation are shown in Fig. 5(f) and Fig. 5(g). For most of the geometries, the spacer selection rule defined above allows to accurately build a reflective bilayer starting from the library of a transmissive single-layer metasurface given the low phase error reported in the color map of Fig. 5(i). The latter represents the phase difference between the FDTD simulations and the analytical product of Eq. (2). The average absolute phase error in this case is 5.3°. The cases that show larger errors are the ones composed of a top nanofin that is much larger than the the bottom one. This can be due to the reflections that occur between the mirror and the base of the top nanofin. This is reminiscent of the observation we made for the case of transmissive bilayer metasurfaces.

## 3.  Conclusion

We showed that a bilayer dielectric metasurface operating in transmission can be expressed as the product of two decoupled single-layer metasurfaces under some constraints. In this process, we distinguished regions in which the bilayer coupling is governed by resonance versus back reflections and we provided systematic recipes to avoid operation in both regimes. Furthermore, we demonstrated that it is also possible to express a reflective bilayer metasurface as the product of five matrices which describe the nanofins composing the structure, the reflective mirror, and a matching spacer in between. By combining our near and far-field analysis, we narrow down the design space to a smaller subset of geometries that are essentially decoupled. Notably by excluding the meta-atoms with strong coupling from the design library—either by avoiding resonant structures or bilayer geometries with very large top nanofins—one can efficiently build a multi-layer metasurface as a cascade of single-layer meta-atoms. In this case, fitting a target profile to a library will entail decomposing it into a product of two matrices and fitting each one following the same selection criteria of a single-layer nanofin. We validated the applicability of our approach to a wide range of libraries by considering titanium dioxide platform at a design wavelength of 532 nm in addition to silicon at 1550 nm, showing the generality of our approach.

## Acknowledgment

We thank Dr. D. Lim, N. Rubin, A. Zaidi, and L. Li, all from Harvard university, for the insightful discussions. The authors from Harvard University acknowledge financial support from Corning Incorporated. Lastly, financial support from the Office of Naval Research (ONR) MURI program, under grant no. N00014-20-1-2450 is acknowledged.

## Disclosures

The authors declare no conflicts of interest.

## Supplemental document

See Supplement 1 for supporting content.

## Data availability

Data underlying the results presented in this paper are not publicly available at this time but may be obtained from the authors upon reasonable request.

# References

1. N. Yu, P. Genevet, M. A. Kats, F. Aieta, J.-P. Tetienne, F. Capasso, and Z. Gaburro, "Light propagation with phase discontinuities: Generalized laws of reflection and refraction," Science **334**, 333–337 (2011).
2. A. H. Dorrah and F. Capasso, "Tunable structured light with flat optics," Science **376**, eabi6860 (2022).
3. S. M. Kamali, E. Arbabi, A. Arbabi, and A. Faraon, "A review of dielectric optical metasurfaces for wavefront control," Nanophotonics **7**, 1041–1068 (2018).
4. P. Genevet, F. Capasso, F. Aieta, M. Khorasaninejad, and R. Devlin, "Recent advances in planar optics: from plasmonic to dielectric metasurfaces," Optica **4**, 139–152 (2017).
5. A. V. Kildishev, A. Boltasseva, and V. M. Shalaev, "Planar photonics with metasurfaces," Science **339**, 1232009 (2013).
6. N. Yu and F. Capasso, "Flat optics with designer metasurfaces," Nat. Mater. **13**, 139–150 (2014).
7. D. Lin, P. Fan, E. Hasman, and M. L. Brongersma, "Dielectric gradient metasurface optical elements," Science **345**, 298–302 (2014).
8. L. Wang, Y. Zhang, X. Guo, T. Chen, H. Liang, X. Hao, X. Hou, W. Kou, Y. Zhao, T. Zhou, S. Liang, and Z. Yang, "A review of THz modulators with dynamic tunable metasurfaces," Nanomaterials **9**, 965 (2019).
9. K. Shastri and F. Monticone, "Nonlocal flat optics," Nat. Photonics **17**, 36–47 (2023).
10. A. Arbabi, Y. Horie, M. Bagheri, and A. Faraon, "Dielectric metasurfaces for complete control of phase and polarization with subwavelength spatial resolution and high transmission," Nat. Nanotechnol. **10**, 937–943 (2015).
11. J. P. Balthasar Mueller, N. A. Rubin, R. C. Devlin, B. Groever, and F. Capasso, "Metasurface polarization optics: Independent phase control of arbitrary orthogonal states of polarization," Phys. Rev. Lett. **118**, 113901 (2017).
12. N. A. Rubin, G. D'Aversa, P. Chevalier, Z. Shi, W. T. Chen, and F. Capasso, "Matrix fourier optics enables a compact full-stokes polarization camera," Science **365**, eaax1839 (2019).
13. E. Arbabi, S. M. Kamali, A. Arbabi, and A. Faraon, "Full-stokes imaging polarimetry using dielectric metasurfaces," ACS Photonics **5**, 3132–3140 (2018).
14. Y. Hu, X. Wang, X. Luo, X. Ou, L. Li, Y. Chen, P. Yang, S. Wang, and H. Duan, "All-dielectric metasurfaces for polarization manipulation: principles and emerging applications," Nanophotonics **9**, 3755–3780 (2020).
15. N. A. Rubin, Z. Shi, and F. Capasso, "Polarization in diffractive optics and metasurfaces," Adv. Opt. Photon. **13**, 836–970 (2021).
16. M. Liu, P. Huo, W. Zhu, C. Zhang, S. Zhang, M. Song, S. Zhang, Q. Zhou, L. Chen, H. J. Lezec, A. Agrawal, Y. Lu, and T. Xu, "Broadband generation of perfect poincaré beams via dielectric spin-multiplexed metasurface," Nat. Commun. **12**, 2230 (2021).
17. Q. Song, X. Liu, C.-W. Qiu, and P. Genevet, "Vectorial metasurface holography," Appl. Phys. Rev. **9**, 011311 (2022).
18. Y. Zhao and A. Alù, "Manipulating light polarization with ultrathin plasmonic metasurfaces," Phys. Rev. B **84**, 205428 (2011).
19. F. Aieta, P. Genevet, M. A. Kats, N. Yu, R. Blanchard, Z. Gaburro, and F. Capasso, "Aberration-free ultrathin flat lenses and axicons at telecom wavelengths based on plasmonic metasurfaces," Nano Lett. **12**, 4932–4936 (2012).
20. A. Pors and S. I. Bozhevolnyi, "Plasmonic metasurfaces for efficient phase control in reflection," Opt. Express **21**, 27438–27451 (2013).
21. J. Li, S. Chen, H. Yang, J. Li, P. Yu, H. Cheng, C. Gu, H.-T. Chen, and J. Tian, "Simultaneous control of light polarization and phase distributions using plasmonic metasurfaces," Adv. Funct. Mater. **25**, 704–710 (2015).
22. J. Park, J.-H. Kang, S. J. Kim, X. Liu, and M. L. Brongersma, "Dynamic reflection phase and polarization control in metasurfaces," Nano Lett. **17**, 407–413 (2017).
23. E. Rahimi and R. Gordon, "Nonlinear plasmonic metasurfaces," Adv. Opt. Mater. **6**, 1800274 (2018).
24. M. S. Bin-Alam, O. Reshef, Y. Mamchur, M. Z. Alam, G. Carlow, J. Upham, B. T. Sullivan, J.-M. Ménard, M. J. Huttunen, R. W. Boyd, and K. Dolgaleva, "Ultra-high-q resonances in plasmonic metasurfaces," Nat. Commun. **12**, 974 (2021).
25. Z. Shi, N. A. Rubin, J.-S. Park, and F. Capasso, "Nonseparable polarization wavefront transformation," Phys. Rev. Lett. **129**, 167403 (2022).
26. Z. Shi, A. Y. Zhu, Z. Li, Y.-W. Huang, W. T. Chen, C.-W. Qiu, and F. Capasso, "Continuous angle-tunable birefringence with freeform metasurfaces for arbitrary polarization conversion," Sci. Adv. **6**, eaba3367 (2020).
27. A. H. Dorrah, N. A. Rubin, A. Zaidi, M. Tamagnone, and F. Capasso, "Metasurface optics for on-demand polarization transformations along the optical path," Nat. Photonics **15**, 287–296 (2021).
28. M. Liu, W. Zhu, P. Huo, L. Feng, M. Song, C. Zhang, L. Chen, H. J. Lezec, Y. Lu, A. Agrawal, and T. Xu, "Multifunctional metasurfaces enabled by simultaneous and independent control of phase and amplitude for orthogonal polarization states," Light. Sci. & Appl. **10**, 107 (2021).
29. R. C. Devlin, A. Ambrosio, N. A. Rubin, J. P. B. Mueller, and F. Capasso, "Arbitrary spin-to-orbital angular momentum conversion of light," Science **358**, 896–901 (2017).
30. Q. Fan, M. Liu, C. Zhang, W. Zhu, Y. Wang, P. Lin, F. Yan, L. Chen, H. J. Lezec, Y. Lu, A. Agrawal, and T. Xu, "Independent amplitude control of arbitrary orthogonal states of polarization via dielectric metasurfaces," Phys. Rev. Lett. **125**, 267402 (2020).
31. B. Xiong, Y. Liu, Y. Xu, L. Deng, C.-W. Chen, J.-N. Wang, R. Peng, Y. Lai, Y. Liu, and M. Wang, "Breaking the limitation of polarization multiplexing in optical metasurfaces with engineered noise," Science **379**, 294–299 (2023).

32. T. Chang, J. Jung, S.-H. Nam, H. Kim, J. U. Kim, N. Kim, S. Jeon, M. Heo, and J. Shin, "Universal metasurfaces for complete linear control of coherent light transmission," Adv. Mater. **34**, 2204085 (2022).
33. Y.-W. Huang, N. A. Rubin, A. Ambrosio, Z. Shi, R. C. Devlin, C.-W. Qiu, and F. Capasso, "Versatile total angular momentum generation using cascaded j-plates," Opt. Express **27**, 7469–7484 (2019).
34. B. Mirzapourbeinekalaye, A. McClung, and A. Arbabi, "General lossless polarization and phase transformation using bilayer metasurfaces," Adv. Opt. Mater. **10**, 2102591 (2022).
35. E. W. Wang, T. Phan, S.-J. Yu, S. Dhuey, and J. A. Fan, "Dynamic circular birefringence response with fractured geometric phase metasurface systems," Proc. National Acad. Sci. **119**, e2122085119 (2022).
36. Y. Bao, F. Nan, J. Yan, X. Yang, C.-W. Qiu, and B. Li, "Observation of full-parameter jones matrix in bilayer metasurface," Nat. Commun. **13**, 7550 (2022).
37. H. Zheng, M. He, Y. Zhou, I. I. Kravchenko, J. D. Caldwell, and J. G. Valentine, "Compound meta-optics for complete and loss-less field control," ACS Nano **16**, 15100–15107 (2022).
38. P. Georgi, Q. Wei, B. Sain, C. Schlickriede, Y. Wang, L. Huang, and T. Zentgraf, "Optical secret sharing with cascaded metasurface holography," Sci. Adv. **7**, eabf9718 (2021).
39. D. F. G. Gallagher and T. P. Felici, "Eigenmode expansion methods for simulation of optical propagation in photonics: pros and cons," in *Integrated Optics: Devices, Materials, and Technologies VII,* vol. 4987 Y. S. Sidorin and A. Tervonen, eds., International Society for Optics and Photonics (SPIE, 2003), pp. 69 – 82.
40. D. M. Whittaker and I. S. Culshaw, "Scattering-matrix treatment of patterned multilayer photonic structures," Phys. Rev. B **60**, 2610–2618 (1999).
41. C. Wan and J. Encinar, "Efficient computation of generalized scattering matrix for analyzing multilayered periodic structures," IEEE Trans. on Antennas Propag. **43**, 1233–1242 (1995).
42. R. Hall, R. Mittra, and K. Mitzner, "Analysis of multilayered periodic structures using generalized scattering matrix theory," IEEE Trans. on Antennas Propag. **36**, 511–517 (1988).
43. S. Colburn, A. Zhan, E. Bayati, J. Whitehead, A. Ryou, L. Huang, and A. Majumdar, "Broadband transparent and cmos-compatible flat optics with silicon nitride metasurfaces [invited]," Opt. Mater. Express **8**, 2330–2344 (2018).

# Supplementary Information
# Do dielectric bilayer metasurfaces behave as a stack of decoupled single-layer metasurfaces?

ALFONSO PALMIERI[1,†], AHMED H. DORRAH[1,†], JUN YANG[2], JAEWON OH[1], PAULO DAINESE[2,*], AND FEDERICO CAPASSO[1,*]

[1]Harvard John A. Paulson School of Engineering and Applied Sciences, Harvard University, Cambridge, Massachusetts 02138, USA

[2]Corning Inc., Painted Post, New York 14870, United States

[†]These authors contributed equally

*DaineseP@corning.com; Capasso@seas.harvard.edu*

**Contents**

## 1. Generating a single-layer metasurface library

In this section, we present a systematic strategy for designing a single-layer metasurface. We start by building a metasurface library based on the results and ideas discussed in [1]. To the aim of obtaining a direct relation between the phase shift and the dimension of the nanofins, a library has been generated by performing a parameter sweep with finite-difference time-domain (FDTD) simulations. In a simplified picture, each subwavelength structure can be considered as a truncated waveguide or a low-quality-factor Fabry-Perot resonator. Nanofins with different dimensions (length and width) will induce different confinement of the field impinging on the structure. This confinement provides an effective refractive index that differs along the two polarization components. The FDTD software used is Ansys Lumerical. It solves Maxwell's equations on a discrete spatial and temporal grid in complex geometries such as the analyzed case. A model of the simulated structure is the one shown in Fig.1(a); it is composed of a $TiO_2$ rectangular nanofin on a fused silica substrate. The boundary conditions applied at the edges of the simulation box are the Periodic Boundary Conditions (PBC) which emulate the existence of an infinitely periodic array of these rectangular fins so that the simulated structure is an infinite array of the same metasurface unit cell.

The dimension of the meta-atom, namely $D_x$ and $D_y$, range from 50 to 250 nm spanning a total of 2601 geometries which cover the phase range from 0 to $2\pi$. The center-to-center separation $d$, which defines the unit cell size, is 420 nm. The structure is illuminated by an x-polarized plane wave source at 532 nm that has been placed at a distance of 600 nm from the bottom of the substrate. At this wavelength the refractive index of the fused silica and of $TiO_2$ are set to 1.46 and 2.48, respectively. For the reflective metasurfaces, we set the refractive index of the Aluminum mirror to $0.811 + i0.366$. A monitor is placed a few wavelengths above the nanofin in order to evaluate the phase response in the far-field and the percentage of transmitted power. The mesh size of this simulation is set to 2.5 nm $\times$ 2.5 nm $\times$ 2.5 nm.

The obtained results are shown in Fig. 1 showing an excellent agreement with those reported in Ref [1]. Note that for the case of a $y$-polarized source, the plots of the phase ($\phi_y$) and power transmission ($T_y$) would be identical to those shown in Fig. 1 but with the $x$ and $y$ axes exchanged (due to symmetry) so they are not shown. It is observed that the considered parameter space is able to completely cover the $0-2\pi$ phase range. The Power Transmission plot depicted in Fig. 1(b) shows that shorter nanofins provide almost uniform unitary transmission while a few geometries with larger $D_x$ manifest resonances and transmission dips. In the complex transmission plot shown in Fig. 1(c) the red dots correspond to the electric field amplitude of each simulated geometry. The average transmission is around 0.91 (black curve) and the red circle represent the unitary circle. Thanks to the obtained results it is possible to map each nanofin geometry to a different (symmetric) Jones matrix.

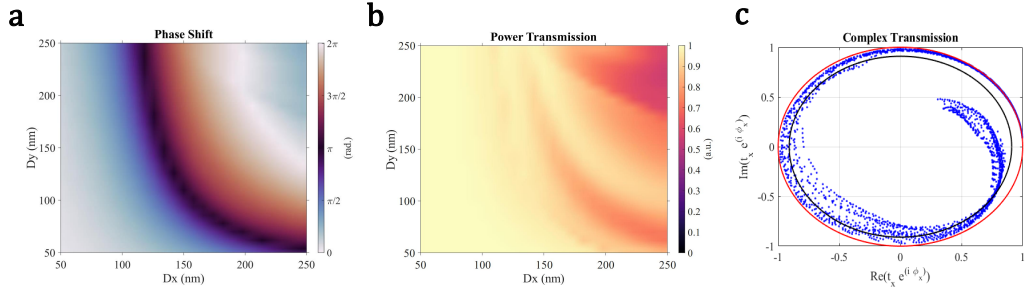

Figure 1. Simulation data for two-dimensional parameter sweeps of $TiO_2$ rectangular fins ($h$ = 600 nm). **(a)** Phase shift $\phi_x$ on x-polarized light. The phase shift has been computed as the ratio between the phase collected in the center of the far field projection of a monitor above the structure when the nanofin is present on top of the substrate and the phase at the same monitor when only the silica substrate is present. Units in radians. **(b)** Power Transmission $T_x$ for x-polarized light. The total power passing through a monitor above the structure relative to the source. **(c)** Complex Transmission. The blue dots represent the electric field plotted on the complex plane for all the simulated geometries. The black circle corresponds to the averaged transmission. The red circle is the unit circle.

## 2. Bilayer simulations: the effect of relative rotation

In this section, we consider a bilayer metasurface and we analyze the effect of relative rotation between its top and bottom nanofins. Two cases are tested: in the first case we set the dimensions of both the bottom and top nanofins to be 134 nm × 202 nm while in the second case we set the dimensions of both nanofins to be 114 nm × 154 nm. The rotation angle of the bottom nanofin around its geometrical center is set to zero ($\theta$ = 0) and we rotate the top nanofin ($\theta'$) between 0° and 90° with increments of 15°. Other rotation angles for the top nanofin beyond $\theta' > 90$ are naturally covered by this sweep (following a simple symmetry argument). In Tables 1 – 4, we report element-by-element comparison between the Jones matrix of the simulated bilayer ($\mathbf{J_s}$) and the "analytical Jones matrix ($\mathbf{J_a}$)" obtained as follows:

$$\mathbf{J_a} = \mathbf{J}_{\text{top}} \cdot \mathbf{J}_{\text{bottom}} = R(-\theta') \cdot \begin{bmatrix} e^{i\phi_{x,\text{top}}} & 0 \\ 0 & e^{i\phi_{y,\text{top}}} \end{bmatrix} \cdot R(\theta') \cdot \begin{bmatrix} e^{i\phi_{x,\text{bottom}}} & 0 \\ 0 & e^{i\phi_{y,\text{bottom}}} \end{bmatrix}, \quad (1)$$

where $R(\theta')$ is the $2 \times 2$ rotation matrix.

|  | $|J_{1,1;a}| - |J_{1,1;s}|$ | $|J_{1,2;a}| - |J_{1,2;s}|$ | $|J_{2,1;a}| - |J_{2,1;s}|$ | $|J_{2,2;a}| - |J_{2,2;s}|$ |
|---|---|---|---|---|
| $\theta' = 0°$ | 0.0653 | 0 | 0 | -0.0548 |
| $\theta' = 15°$ | 0.0307 | -0.0154 | 0.0434 | -0.0547 |
| $\theta' = 30°$ | -0.0203 | -0.0345 | 0.0148 | -0.0386 |
| $\theta' = 45°$ | -0.0281 | -0.0404 | -0.0280 | -0.0155 |
| $\theta' = 60°$ | -0.0398 | -0.0381 | -0.0443 | -0.0101 |
| $\theta' = 75°$ | -0.0643 | -0.0257 | -0.0328 | -0.0225 |
| $\theta' = 90°$ | -0.0776 | 0 | 0 | -0.0323 |

Table 1. Element-by-element comparison between the magnitude of the Jones matrix elements of the simulated bilayer $\mathbf{J_s}$ and the elements of the analytical Jones matrix $\mathbf{J_a}$ for the bilayer composed of two nanofins of dimensions 134 nm × 202 nm.

|  | $\angle(J_{1,1;a}) - \angle(J_{1,1;s})$ | $\angle(J_{1,2;a}) - \angle(J_{1,2;s})$ | $\angle(J_{2,1;a}) - \angle(J_{2,1;s})$ | $\angle(J_{2,2;a}) - \angle(J_{2,2;s})$ |
|---|---|---|---|---|
| $\theta' = 0°$ | 8.6° | 0 | 0 | 2.5° |
| $\theta' = 15°$ | 6.7° | 6.5° | 9.0° | 3.5° |
| $\theta' = 30°$ | 6.3° | 6.6° | 4.4° | 3.5° |
| $\theta' = 45°$ | 7.3° | 7.0° | 4.5° | 3.9° |
| $\theta' = 60°$ | 5.1° | 6.5° | 4.4° | -4.8° |
| $\theta' = 75°$ | 4.4° | 6.3° | 4.2° | 5.4° |
| $\theta' = 90°$ | 3.2° | 0 | 0 | 5.1° |

Table 2. Element-by-element comparison of the phase of the Jones matrix elements of the simulated bilayer $\mathbf{J_s}$ and the elements of the analytical Jones matrix $\mathbf{J_a}$ for the bilayer composed of two nanofins of dimensions 134 nm × 202 nm.

|  | $|J_{1,1;a}| - |J_{1,1;s}|$ | $|J_{1,2;a}| - |J_{1,2;s}|$ | $|J_{2,1;a}| - |J_{2,1;s}|$ | $|J_{2,2;a}| - |J_{2,2;s}|$ |
|---|---|---|---|---|
| $\theta' = 0°$ | -0.0291 | 0 | 0 | 0.0112 |
| $\theta' = 15°$ | -0.0343 | 0.0277 | 0.0122 | 0.0564 |
| $\theta' = 30°$ | -0.0324 | 0.0174 | -0.0036 | 0.0375 |
| $\theta' = 45°$ | -0.0367 | 0.0029 | 0.0085 | 0.0025 |
| $\theta' = 60°$ | -0.0523 | -0.0002 | -0.0005 | -0.0309 |
| $\theta' = 75°$ | -0.0609 | 0.0108 | 0.0158 | -0.0469 |
| $\theta' = 90°$ | -0.0560 | 0 | 0 | -0.0439 |

Table 3. Element-by-element comparison of the magnitude of the Jones matrix elements of the simulated bilayer $\mathbf{J_s}$ and the elements of the analytical Jones matrix $\mathbf{J_a}$ for the bilayer composed of two nanofins of dimensions 114 nm × 154 nm.

|  | $\angle(J_{1,1;a}) - \angle(J_{1,1;s})$ | $\angle(J_{1,2;a}) - \angle(J_{1,2;s})$ | $\angle(J_{2,1;a}) - \angle(J_{2,1;s})$ | $\angle(J_{2,2;a}) - \angle(J_{2,2;s})$ |
|---|---|---|---|---|
| $\theta' = 0°$ | -1.8° | 0 | 0 | 1.7° |
| $\theta' = 15°$ | -1.2° | -3.9° | 1.6° | 0.8° |
| $\theta' = 30°$ | -1.2° | 1.7° | 0.1° | 1.2° |
| $\theta' = 45°$ | -1.6° | 2.2° | 0.7° | 0.1° |
| $\theta' = 60°$ | -1.7° | 0.2° | 1.8° | 1.3° |
| $\theta' = 75°$ | -2.6° | 2.6° | 3.4° | 2.7° |
| $\theta' = 90°$ | 2.2° | 0 | 0 | 2.6° |

Table 4. Element-by-element comparison of the phase of the Jones matrix elements of the simulated bilayer $\mathbf{J_s}$ and the elements of the analytical Jones matrix $\mathbf{J_a}$ for the bilayer composed of two nanofins of dimensions 114 nm × 154 nm.

The discrepancy that arises between the Jones matrix of the simulated bilayer and the "analytical

Jones matrix" is on average not so large and can be considered acceptable for both the case of phase and amplitude. This implies that, at least for the tested geometries, the output response can be calculated using Eq. (5). To reconcile the discrepancies, recall that we apply the Periodic Boundary Conditions (PBC) when simulating both the single and bilayer metasurfaces. The PBC emulates a periodic array composed of identical nanofins with zero rotation angle everywhere producing an electric field at the borders of the unit cell that is different from the case of a rotated nanofin. Although the nanofins are rotated, the square unit cells comprising the simulated structure are always fixed. Hence, the lattice symmetry is different from the case reported in Fig. 1. Accordingly, applying the rotation matrix on the simulation data of the aligned single-layer nanofins—calculating Eq. (5)—does not entirely capture their response under rotation and is thus expected to slightly differ from the full-wave simulations.

The data reported in Tables 1–4 suggest that the discrepancy between the analytical and simulated cases is higher for the bilayer geometry of dimensions 134 nm × 202 nm. This is likely due to the larger mismatch between the major and minor axes (compared to the 114 nm × 154 nm geometry) which in turn reduces the overlap region between the top and bottom nanofins, thereby introducing more reflections and making this geometry more sensitive to the rotation-dependent variation of the electric field at the boundary.

## 3. Phase coverage of the bilayer metasurface operating in transmission

Here, we report the results obtained from the simulation of the structure shown in Fig. 1(b). The dimension of the bottom nanofin are fixed (134 nm × 202 nm) while the dimensions of the top nanofin are swept between 50 nm and 250 nm. Figure 2 depicts the phase shift of the diagonal elements of the Jones matrix of the simulated structure, $J_{1,1}$ and $J_{2,2}$. These results suggest that both elements (in the analyzed parameter space) can cover a phase range between 0 and $2\pi$ as reported on the vertical axis of the plots. The color bars depict the error between the simulated results and the analytical ones for both Jones matrix elements. Similar results were obtained for the case of a bilayer metasurfce operating in reflection with fixed bottom nanofin dimensions.

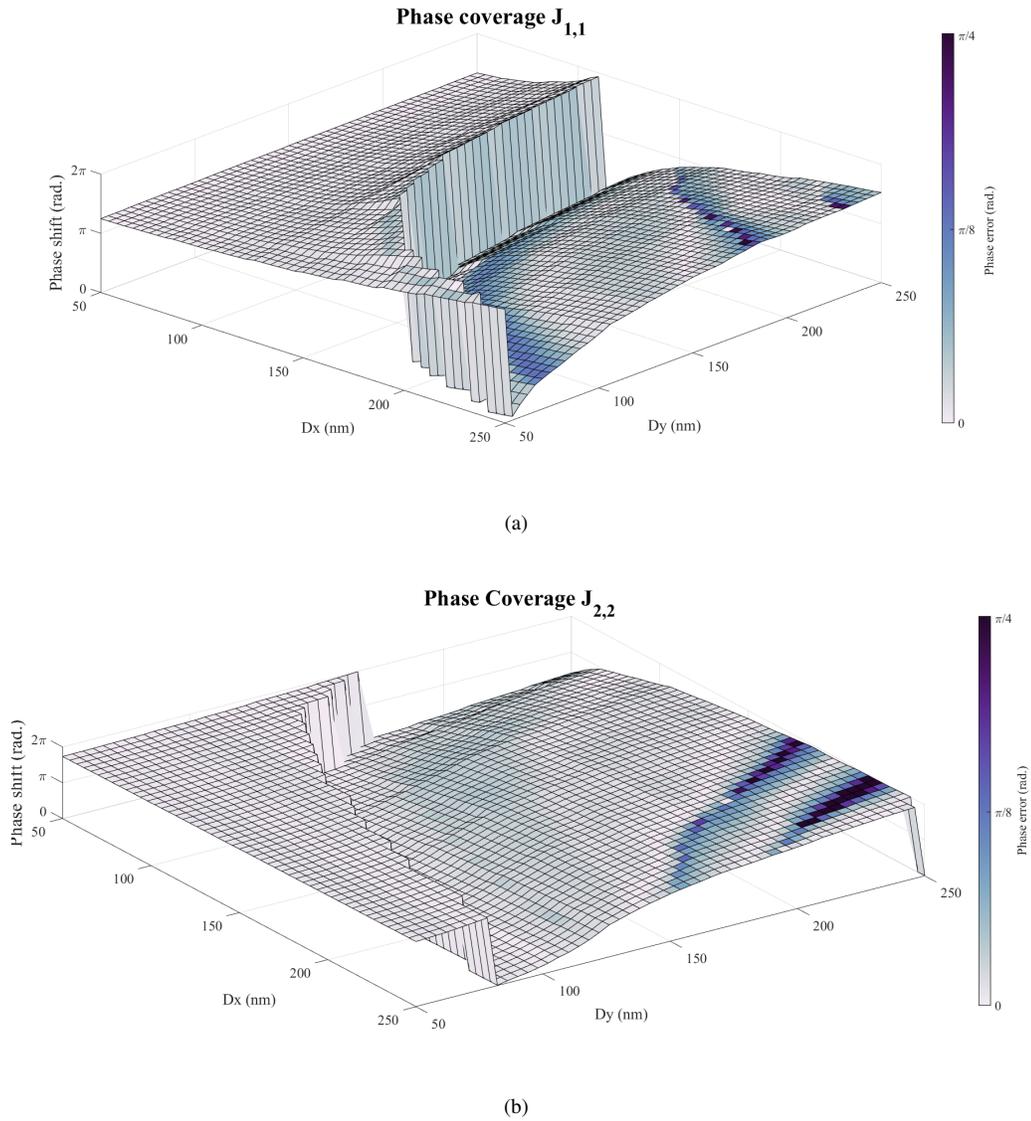

Figure 2. Phase error between the Jones matrix elements of the full-wave simulated bilayer and the corresponding analytically computed ones. Full 0 to $2\pi$ phase coverage can be achieved with low phase error. Here the bottom nanofin dimensions are fixed. (a) Difference between the phase of element $J_{1,1}$ extracted from the simulation and the phase of $J_{1,1}$ analytically computed from the data of the single layer. The vertical axis depicts the phase shift of each geometry, suggesting 0 to $2\pi$ phase coverage. (b) Difference between the magnitude of $J_{2,2}$ extracted from the simulation and analytical calculations. Full phase coverage can again be achieved, as before.

## 4. The effect of resonance and of back reflections

There are two main sources of errors when describing the bilayer as a stack of two decoupled single layers: a) Fabry-Perot-like back reflections that typically occur between the top nanofin and the substrate if the top nanofin is larger than the bottom one, and b) coupling through the evanescent fields when either nanofin operates near resonance. In this section, we study these sources of error in more detail. Figure 3 shows the field profile in the xz-plane of all possible geometries that cover these two sources of error. The results were obtained using full wave (FDTD) simulations. By varying the gap between the top and bottom nanofins and recording the amplitude response, we gain some insights into the bilayer coupling strength.

In Fig. 3(a) we consider a bilayer meta-atom composed of two nanofins at resonance. In this case, as it has been pointed out in the near-field section, the two nanofins are coupled through the evanescent modes (which are more significant when the nanofins are in proximity). Consequently, by varying the gap between the two nanofins, the field profiles and their confinement is altered. Such behavior is consistent with the S- and T- matrix analysis (of Fig. 3(c)) where a discrepancy between full S- and T-matrices versus the $(2 \times 2)$ matrices was observed.

Figure 3(b) depicts another case in which the top nanofin is at resonance and is larger than the bottom one. Interference fringes are observed between the top nanofin and the substrate due to the back reflections. This is consistent with the oscillations in transmission amplitude observed in Fig.3(b). The mode shape in the top nanofin is unperturbed. Hence, the effect of back reflections is more significant than resonance, causing the two nanofins to be strongly coupled. To complement this picture, we consider another geometry in which the top nanofin is still at resonance but is smaller than the nanofin bottom; hence, the back reflections are mitigated. In this case, the field profile still does not change as a function of gap size and no interference fringes are observed. Therefore, unlike the previous case, the two nanofins can be considered decoupled even in the presence of resonance.

Resonance becomes much more significant when it occurs in the bottom nanofin; the bilayer coupling becomes strong due to the interaction of the evanescent fields. If the top nanofin is too large (considering our parameter space, if its dimensions roughly exceed 130 nm $\times$ 130 nm), these evanescent fields are strongly coupled. This is confirmed by the FDTD simulations shown in Fig. 3(d) where the mode profiles in both the top and bottom nanofins are clearly perturbed as the gap size is varied.

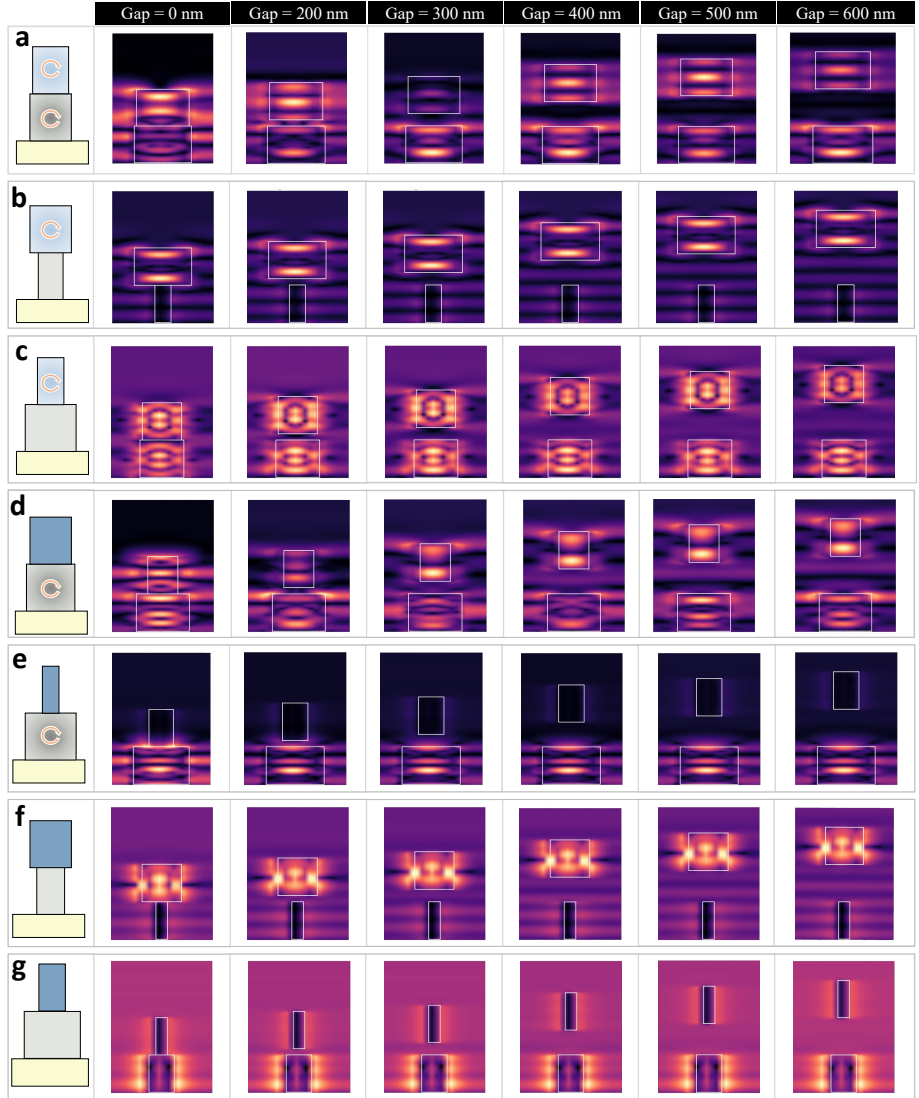

Figure 3. Simulated field profiles of different examples bilayer meta-atoms (in the xz-plane). The input polarization is $E_x$ (in plane). **(a)** Both top and bottom nanofin are at resonance; exhibiting strong coupling. **(b)** Top nanofin is at resonance and is larger than the bottom one; causing back reflections. **(c)** Top nanofin is at resonance but is smaller in size than the bottom one; exhibiting very weak coupling. **(d)** Bottom nanofin is at resonance. The top nanofin dimensions are fairly large (170 nm × 170 nm); showing strong coupling. **(e)** Bottom nanofin at resonance. The top nanofin dimensions are fairly small (90 nm × 90 nm); effectively decoupled. **(f)** Neither bottom nor top nanofin at resonance. Top nanofin is larger than the bottom one; introducing back reflections. **(g)** Neither the bottom nor the top nanofins are at resonance. Top nanofin is smaller than the bottom one; a perfectly decoupled geometry.

On the other hand, in the limit when the top nanofin is much smaller (while the bottom is still at resonance), the coupling is weak and the field profile does not change by varying the gap size as shown in Fig. 3(e).

Figure 3(f) emphasizes the effect of back reflections in the absence of resonance (both nanofins are off-resonance). One can observe the Fabry-Perot-like interference fringes that arise between the top nanofin and the substrate. This effect was not captured by the $(2 \times 2)$ T-matrix but rather by the S- matrices. This confirms that this geometry cannot be treated as a stack of two decoupled single layers.

Lastly, in Fig. 3(g) we consider a more straight forward case with neither resonances nor back reflections. Here, both nanofins operate off-resonance. The top nanofin is smaller than the bottom one. As it can be noticed, the field profile remains identical regardless of the gap size. No fringes arise and the two nanofins are effectively decoupled.

## 5. Single-layer metasurface in reflection

Here, we show the simulation results of the single-layer metasurface operating in reflection without introducing a spacer between the mirror and the nanofin. As mentioned in the main text, looking at the results of the analytical calculations (Fig. 4(a)) and those of the FDTD simulations (Fig. 4(b)), one can immediately notice the significant differences over the full parameter space. The average phase error between the two cases is on the order of $10°$.

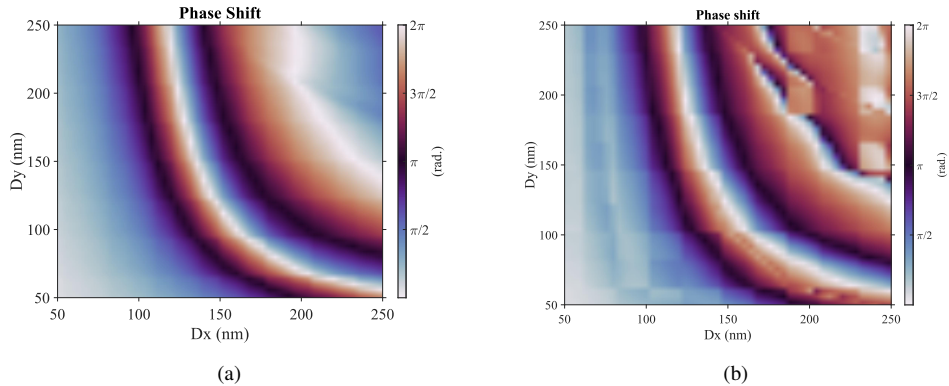

Figure 4. single-layer metasurface working in reflection. **(a)** Phase shift obtained from the analytical product. **(b)** Phase shift obtained from FDTD simulation of a single-layer metasurface working in reflection without the introduction of the spacer.

## 6. The effect of a relative rotation of 45° in reflection: sweeping the dimensions of the top nanofin while fixing the dimensions of the bottom one

To verify if the assumption of decoupling is also valid when a relative rotation between the metasurface nanofins is introduced, we simulated a bilayer structure while fixing the bottom nanofin and sweeping the dimension of the top nanofin. Here, the top nanofin is rotated with respect to the bottom one by an angle of 45°. In this case it is necessary to introduce a rotation matrix $R(\theta)$ that sandwiches the Jones matrix of the top nanofin to take into account the rotation of the top nanofin. The Jones matrix of this structure can be analytically written as follows:

$$\mathbf{J_a} = \mathbf{J}_{\text{bottom}} \cdot \mathbf{J}_{\text{top}} = R(-\theta') \cdot \begin{bmatrix} e^{i\phi_{x,\text{bottom}}} & 0 \\ 0 & e^{i\phi_{y,\text{bottom}}} \end{bmatrix} \cdot R(\theta') \cdot \begin{bmatrix} e^{i\phi_{x,\text{top}}} & 0 \\ 0 & e^{i\phi_{y,\text{top}}} \end{bmatrix}, \qquad (2)$$

where $R(\theta)$ is the $2 \times 2$ rotation matrix that is defined as:

$$R(\theta) = \begin{bmatrix} \cos\theta & -\sin\theta \\ \sin\theta & \cos\theta \end{bmatrix}. \qquad (3)$$

Hence, by introducing a relative rotation, it is possible to obtain a metasurface where the four Jones matrix elements are all different. In Supplementary Figures 5 and 6 we show the magnitude and phases of the four elements of the Jones matrix describing the bilayer obtained via FDTD simulations. Interestingly, the geometries that now exhibit larger phase error — for elements $J_{1,1}$ and $J_{2,2}$ — are not the ones in the top right of the plot (as was the case with no rotation) but rather the ones in the top left and bottom right.

The average phase error is 1.14° for the element $J_{1,1}$ and 1.02° for the element $J_{2,2}$, exhibiting less error compared to the case where no rotation angle was introduced. This is because the rotation increases the overlap region between the top and bottom nanofins, thereby reducing multiple reflections between the top nanofin and the substrate. For the off diagonal Jones matrix elements, the largest phase error is in both cases on the diagonal of the plot. On the other hand, the phase of these element is not relevant since, as it can be noticed from Fig. 5, the magnitude of these elements is zero. The average absolute phase error for the elements $J_{1,2}$ and $J_{2,1}$, excluding the points on the diagonal is around 3°.

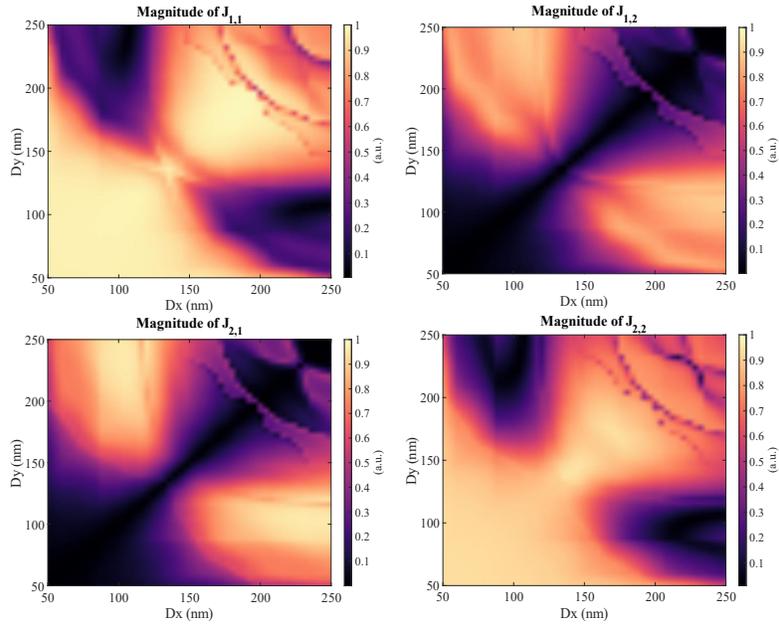

Figure 5. Magnitude of the four elements of the Jones matrix of a bilayer metasurface whose bottom nanofins dimensions are set to be 134 nm × 202 nm while the dimensions of the top nanofin are sweeping between 50 and 250 nm.

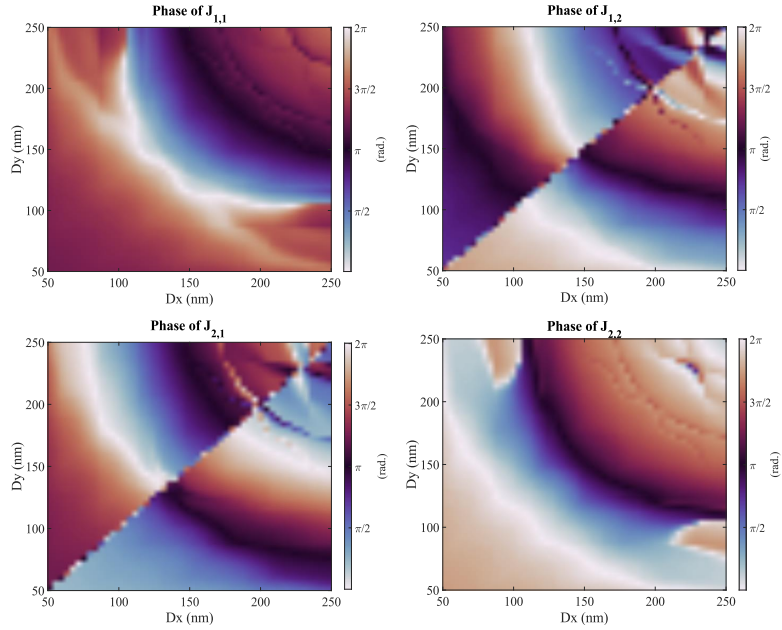

Figure 6. Phase of the four elements of the Jones matrix of a bilayer metasurface whose bottom nanofins dimensions are set to be 134 nm × 202 nm while the dimensions of the top nanofin are sweeping between 50 and 250 nm.

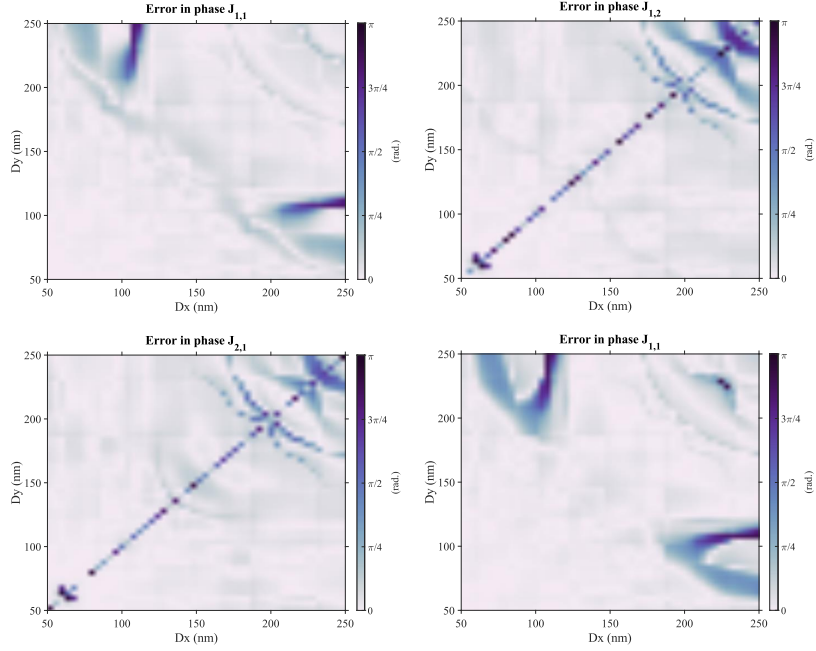

Figure 7. Error in phase of the four elements of the Jones matrix of a bilayer metasurface whose bottom nanofins dimensions are set to be 134 nm × 202 nm while the dimensions of the top nanofin are sweeping between 50 and 250 nm.

## 7. Near-field analysis for a Silicon bilayer metasurface

In this section, we expand our analysis by considering silicon as another material platform at a design wavelength, $\lambda = 1550$ nm. Our aim is to show that the physical dynamics associated with back reflections and coupling in a bilayer metasurface are, to some extent, independent of the material platform, its parameter space, and wavelength. The phase and transmission library for a single layer silicon nanofin is depicted in Fig. 8. Using this library, we then performed a near-field analysis for a silicon bilayer metasurface at 1550 nm, considering a unit-cell size of $500 \times 500$ nm and 1-$\mu$m tall nanofins.

We adopted the same scattering and transmission matrix approaches used for the TiO$_2$ platform in the main text. We tested the same four categories considered before, assuming x-polarized incident light. Figure 9(a) shows the first case: a bilayer metasurface composed of two off-resonance nanofins with a smaller nanofin at the top. It is observed that the Full S-matrix transmission oscillates slightly (0.01%) around a mean value which matches the (2x2) T-matrix. In this case, the Jones matrix is able to correctly describe the nanofin dynamics as if they were decoupled. This is consistent with the results obtained for TiO$_2$. Furthermore, in the same figure, we show the field profile in the xz-plane obtained using FDTD simulations. In this case, since neither the evanescent coupling nor back reflections play a significant role, the field profile remains identical regardless of the gap size. Figure 9(b) shows the case when the top nanofin is larger than the bottom one. Due to the size mismatch between the bottom and top nanofins, the effect of back reflections result in a large deviation between the responses of the S-matrices and the T-matrices. More specifically, the Fabry-Perot cavity effect created between the top nanofin and the substrate is emphasized here by looking at the field profile and the associated interference fringes. Figure 9(c) represents the third case where two identical nanofins operate at resonance. Here, the coupling trough evanescent modes causes a large discrepancy between the Full S- (and T-) matrices, compared to the (2x2) T (and S) matrices. By looking at the field profiles, we observe that the field confinement due to resonance is highly dependent on the gap size. On the other hand, if the top nanofin is much smaller, while the bottom is still at resonance, the evanescent coupling becomes weak and field confinement exhibit less dependence on the gap size. This is depicted in 9(d) where the full S-matrix and the (2x2) T-matrix responses are in very good agreement. In summary, this analysis shows that the rules governing the evanescent coupling and back reflections in bilayer metasurfaces are somewhat universal, regardless of the material platform and design wavelength. The underlying physics dictating if a bilayer meta-atom can be described as two decoupled nanofins thus apply to different metasurface libraries.

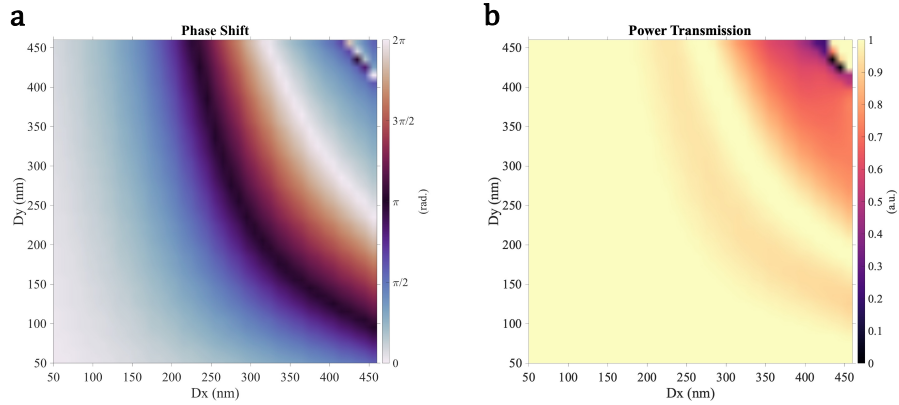

Figure 8. Simulation data for two-dimensional parameter sweeps of Si rectangular fins ($h = 1000$ nm). **(a)** Power Transmission $T_x$ for x-polarized light. The total power passing through a monitor above the structure relative to the source. **(b)** Phase shift $\phi_x$ on x-polarized light. The phase shift has been computed as the ratio between the phase collected in the center of the far field projection of a monitor above the structure when the nanofin is present on top of the substrate and the phase at the same monitor when only the silica substrate is present. Units in radians.

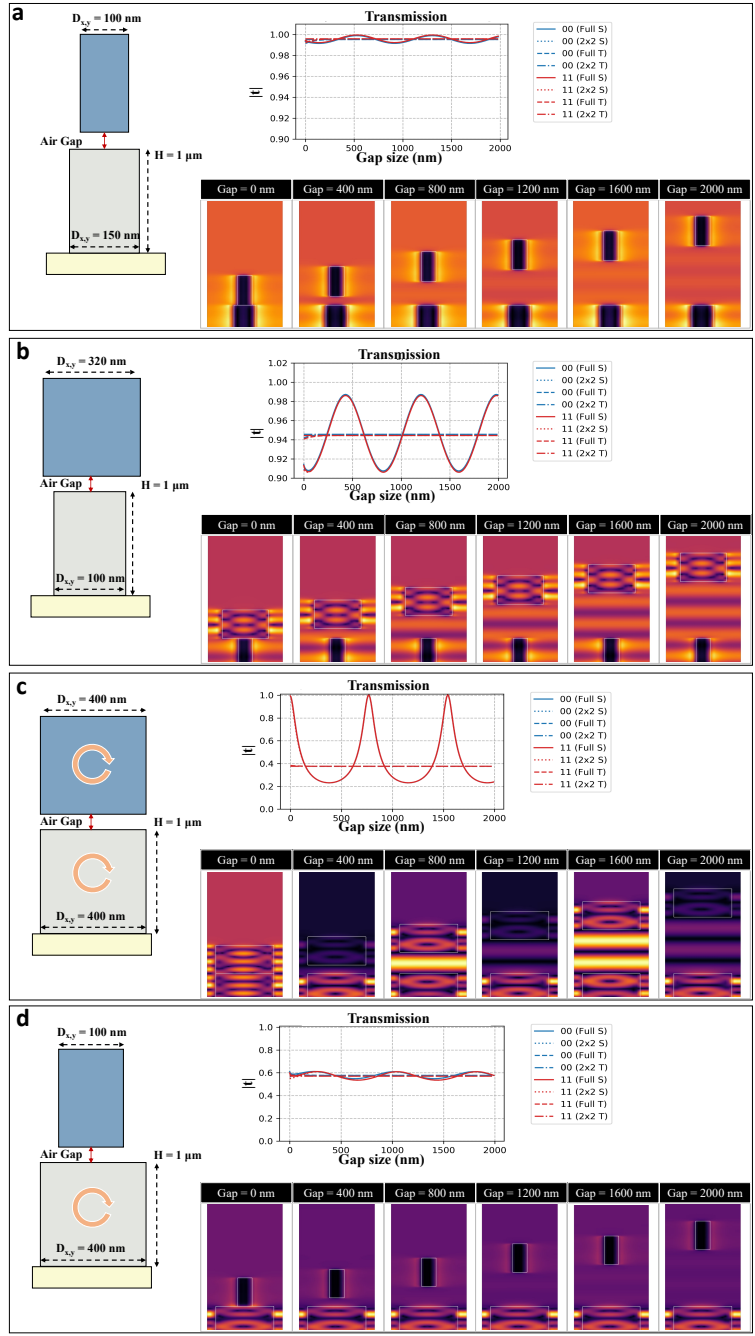

Figure 9. Transmission amplitude response from S- and T-matrices under x-polarization and FDTD profiles for a Silicon bilayer metasurface operating at 1550nm. Four cases are considered. (a) Two nanofins operating off resonance. (b) Bilayer meta-atom composed of a top nanofin larger than the bottom. (c) Two identical nanofins operating at resonance (d) Only bottom nanofin of the bilayer meta-atom is at resonance and is larger than the top nanofin.

## References

1. J. P. Balthasar Mueller, N. A. Rubin, R. C. Devlin, B. Groever, and F. Capasso, "Metasurface polarization optics: Independent phase control of arbitrary orthogonal states of polarization," Phys. Rev. Lett. **118**, 113901 (2017).